\documentclass[english]{article}
\usepackage[T1]{fontenc}
\usepackage[latin9]{inputenc}
\usepackage{amsmath}
\usepackage{graphicx}

\makeatletter

\providecommand{\tabularnewline}{\\}

\usepackage{spconf,amsmath,graphicx,url}
\usepackage[pdfpagemode=UseNone,pdfstartview=FitH,pdfview=FitH,colorlinks=true,
 pdftitle=Daala:\ A\ Perceptually-Driven\ Still\ Picture\ Codec,
 pdfauthor=Xiph.Org]{hyperref}

\name{Jean-Marc Valin, Nathan E. Egge, Thomas Daede, Timothy B. Terriberry, Christopher Montgomery
\thanks{Send correspondence to Jean-Marc Valin <\href{mailto:jmvalin@jmvalin.ca}{jmvalin@jmvalin.ca}>.}
}
\address{Mozilla, Mountain View, CA, USA \\ Xiph.Org Foundation}
\title{Daala: A Perceptually-Driven Still Picture Codec}

\makeatother

\usepackage{babel}
\begin{document}
\maketitle
\begin{abstract}
Daala is a new royalty-free video codec based on perceptually-driven
coding techniques. We explore using its keyframe format for still
picture coding and show how it has improved over the past year. We
believe the technology used in Daala could be the basis of an excellent,
royalty-free image format.
\end{abstract}

\section{Introduction}

Daala is a royalty-free video codec designed to avoid traditional
patent-encumbered techniques used in most current video codecs. In
this paper, we propose to use Daala's keyframe format for still picture
coding. In June 2015, Daala was compared to other still picture codecs
at the 2015 Picture Coding Symposium (PCS)~\cite{EggePCS,pcs2015}.
Since then many improvements were made to the bitstream to improve
its quality. These include reduced overlap in the lapped transform,
finer quantization of chroma, encoder search improvements, as well
as a new deringing filter. 

Rather than describe in detail all of the coding techniques used in
Daala, this paper provides references to previously published documents
describing them. We then present improvements since the version presented
at PCS~2015. Finally, we present the results obtained and compare
them to previously reported results~\cite{EggePCS}.

\section{Fundamental Daala Techniques}

One of the goals of the Daala codec is to explore techniques that
are very different from those typically used in most codecs. Most
of these techniques have been fundamental to Daala since the initial
stages of the project. They are described below.

\subsection*{Lapping}

Daala uses lapped transforms~\cite{MalvarS89,Tran2003} rather than
a regular DCT followed by a deblocking filter. This reduces blocking
artifacts but prevents the use of standard pixel-based intra-prediction
techniques~\cite{DaedeDCC}. Instead, we use a simple frequency-domain
intra predictor that only handles horizontal and vertical directions~\cite{EggePCS}.
Also, DC coefficients are combined recursively using a Haar transform,
up to the level of 64x64 superblocks.

\subsection*{Multi-Symbol Entropy Coder}

Most recent video codecs encode information using binary arithmetic
coding, meaning that each symbol can only take two values. The Daala
range coder supports up to 16 values per symbol, making it possible
to encode fewer symbols~\cite{derfTools}. This is equivalent to
coding up to four binary values in parallel and reduces serial dependencies.

\subsection*{Perceptual Vector Quantization}

Rather than use scalar quantization like the vast majority of picture
and video codecs, Daala is based on perceptual vector quantization
(PVQ)~\cite{valin2015spie}. PVQ makes it possible to take into account
masking effects with no extra signaling.

\subsection*{Chroma from Luma (CfL) Prediction}

Because of its structure, PVQ makes it especially easy to predict
chroma planes from the luma plane. Daala's chroma from luma (CfL)~\cite{egge2015spie}
prediction uses the luma transform coefficients to predict the chroma
transform coefficients.

\section{Recent Improvements}

\label{sec:recent-improvements}

Since the last evaluation at PCS, Daala has improved significantly.
In addition to general encoder tuning, there have been more visible
changes, each described below.

\subsection{Deringing Filter}

The previous version of Daala evaluated in~\cite{EggePCS} included
a deringing filter based on a directional \emph{paint} algorithm.
While it provided an improvement in quality, the paint deringing filter
proved very hard to vectorize and efficiently implement. This was
mostly due to the complicated per-pixel weights used in block interiors,
which blended up to four points from the block boundary, selected
in ways that were not regular over an entire block. 

In the current version of Daala, the paint deringing filter is replaced
by a new directional deringing filter based on a conditional replacement
filter~\cite{ValinDeringing,DaedeDCC}. Let $x\left(n\right)$ denote
a 1-dimensional signal and $w_{k}$ denote filter tap weights, a linear
finite impulse response (FIR) filter with unit DC response is defined
as
\begin{equation}
y\left(n\right)=\frac{1}{\sum_{k}w_{k}}\sum_{k}w_{k}x\left(n+k\right)\ ,\label{eq:FIR1}
\end{equation}
which can alternatively be written as
\begin{equation}
y\left(n\right)=x\left(n\right)+\frac{1}{\sum_{k}w_{k}}\sum_{k,k\neq0}w_{k}\left[x\left(n+k\right)-x\left(n\right)\right]\ .\label{eq:FIR2}
\end{equation}
The main advantage of expressing a filter in the form of Eq.~(\ref{eq:FIR2})
is that the normalization term $\frac{1}{\sum_{k}w_{k}}$ can be approximated
relatively coarsely without affecting the unit gain for DC. This makes
it easy to use small integers for the weights $w_{k}$.

The disadvantage of linear filters for removing ringing artifacts
is that they tend to also cause blurring. To reduce the amount of
blurring, the conditional replacement filter used in Daala excludes
the signal taps $x\left(n+k\right)$ that would cause blurring and
replaces them with $x\left(n\right)$ instead. This is determined
by whether $x\left(n+k\right)$ differs from $x\left(n\right)$ by
more than a threshold $T$. The FIR filter in Eq.~(\ref{eq:FIR2})
then becomes a conditional replacement filter expressed as
\begin{equation}
y\left(n\right)=x\left(n\right)+\frac{1}{\sum_{k}w_{k}}\sum_{k,k\neq0}w_{k}R\left(x\left(n+k\right)-x\left(n\right),T\right)\ ,\label{eq:CRF}
\end{equation}
where
\begin{equation}
R\left(x,T\right)=\left\{ \begin{array}{ll}
x & ,\ \left|x\right|<T\\
0 & ,\ \mathrm{otherwise}
\end{array}\right.\ .
\end{equation}

To further reduce the risk of blurring the decoded image, the conditional
replacement filter is applied along the main direction of the edges
in each 8x8 block. The algorithm for finding directions is described
in~\cite{ValinDeringing} and is the same as the one previously used
in the paint deringing filter. For each 8x8 block, it determines which
of eight different directions best represents the content of the block.
It can be efficiently implemented in SIMD. A 7-tap conditional replacement
filter is applied in Daala for a single pixel in a 8x8 block. The
process is repeated for each pixel in each block being filtered.

To reduce ringing in very smooth regions of the image, the filter
is applied a second time to combine multiple output values of the
first filter. The second filter is applied either vertically or horizontally
-- in the direction most orthogonal to the one used in the first filter.
For example, for a 45-degree direction, the second filter would be
applied vertically. The combined effect of the two filters is a separable
deringing filter that covers a total of 35~pixel taps.

\begin{figure}
\centering{\includegraphics[width=0.9\columnwidth]{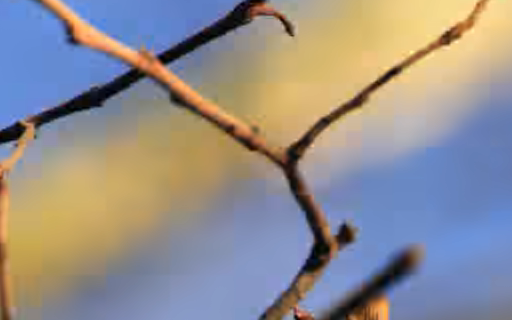}}

\centering{\includegraphics[width=0.9\columnwidth]{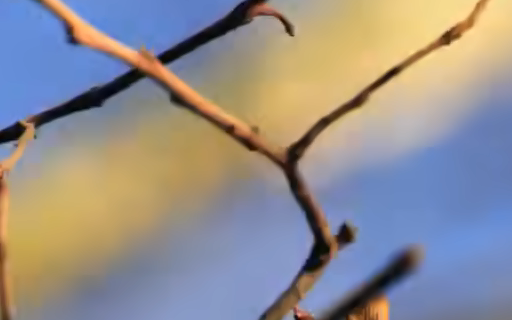}}\caption{Effect of the deringing filter at low bitrate. Top: without deringing.
Bottom: with deringing.\label{fig:Effect-of-deringing}}
\end{figure}

Fig.~\ref{fig:Effect-of-deringing} shows the effect of the deringing
filter on edges at low bitrate. For the impact on objective metrics,
see Table~\ref{tab:bdrate-deringing} in Section~\ref{sec:results}.
An interactive demonstration of the deringing filter is also available~\cite{deringing-demo}.

\subsection{64x64 DCT}

Larger transform sizes generally provide better coding efficiency,
particularly for HD and UHD content. Since the last subjective test~\cite{pcs2015},
Daala has added support for block sizes up to 64x64. When a 64x64
block is used in a keyframe, only the lower quadrant of coefficients
is quantized and coded. This provides additional performance as regions
where 64x64 blocks are used typically lack high frequency detail.
This also makes it possible for a Daala based picture codec to use
a lower complexity 64-point DCT which does not compute the upper 32
frequencies.

\subsection{Reduced Lapping}

Lapped transforms have complex interactions with variable transform
sizes. In a previous version of Daala that used maximum lapping width
for any given transform size, making block size decisions based on
rate-distortion optimization (RDO) proved computationally intractable.
This is why the lapping as implemented in~\cite{EggePCS} was fixed
at 8-point almost everywhere, with the exception that an 8x8 block
being subdivided into four 4x4 blocks used 4-point lapping on the
internal boundaries. This caused 4x4 transform blocks to have 4-point
lapping on two sides and 8-point lapping on the other two sides. The
current version of Daala uses 4-point lapping on all block boundaries.
Although it makes textures slightly worse and causes slightly more
blocking artifacts, the ringing on edges is significantly reduced
by using 4-point lapping.

\subsection{Reduced Overhead Entropy Coder}

The original entropy coder in Daala is based on a multiply-free algorithm
described in~\cite{stuiver1998piecewise}. We have replaced the piecewise
linear integer mapping from that paper with a new mapping that has
less approximation error. Using the notation from~\cite{stuiver1998piecewise},
the new partition function is
\begin{align}
e & =\max(2R-3t,0)\\
f(x,t,R) & =x+\min(x,e)\nonumber \\
 & +\min\Bigl(\Bigl\lfloor\frac{\max(x-e,0)}{2}\Bigr\rfloor,R-t\Bigr)
\end{align}
This mapping is about three times as expensive to evaluate, but remains
an order of magnitude less complex than a division and reduces the
average overhead introduced by entropy coder by about 0.3\%. While
this is an acceptable trade-off for still picture coding, it may ultimately
prove to be too expensive for real-time video to justify this reduction.

\subsection{Finer Chroma Quantization}

\label{sub:finer-chroma}

Among the most visible artifacts from the Daala version in~\cite{EggePCS}
are chroma quantization artifacts caused by the chroma quantization
being coarser than luma quantization at all bitrates. Since then,
the chroma quantizers have changed to be coarser than luma at high
bitrate, but finer at low bitrate. Although this results in a significant
improvement in visual quality, it obviously regresses all the commonly-used
luma-only metrics.

\section{Results}

\label{sec:results}

Because of the changes in chroma quantization (Section~\ref{sub:finer-chroma}),
it is hard to show the recent improvements in Daala using objective
metrics. For this reason, we will show metrics for the current version
\emph{minus} the chroma quantization changes. The PSNR, PSNR-HVS~\cite{PSNRHVSM},
SSIM, and FAST-SSIM~\cite{FastSSIM} metrics are obtained using the
\emph{Are We Compressed Yet?}\footnote{\url{https://arewecompressedyet.com/}}
testing infrastructure. A test set \emph{subset1}\footnote{\url{https://people.xiph.org/~tterribe/daala/subset1-y4m.tar.gz}}
composed of 50~still images is compressed and the metrics averaged
together\footnote{\url{https://github.com/tdaede/rd_tool}}. Table~\ref{tab:bdrate-pcs-badchroma}
shows the recent improvements at low (0.05 to 0.2~bit/pixel), medium
(0.2 to 0.5~bit/pixel), and high (0.5 to 1.0~bit/pixel) bitrate.
Among the improvements listed in Section~\ref{sec:recent-improvements},
the one with the largest impact on metrics is the deringing filter,
as shown in Table~\ref{tab:bdrate-deringing}.

\begin{table}
\centering{%
\begin{tabular}{|c|c|c|c|}
\hline 
Metric & Low (\%) & Medium (\%) & High (\%)\tabularnewline
\hline 
\hline 
PSNR & -6.3 & -6.9 & -7.9\tabularnewline
\hline 
PSNR-HVS & -6.7 & -6.6 & -6.3\tabularnewline
\hline 
SSIM & -4.8 & -5.3 & -6.5\tabularnewline
\hline 
FAST-SSIM & -2.4 & -2.2 & -0.6\tabularnewline
\hline 
\end{tabular}}

\caption{Bjøntegaard-delta~\cite{testing-draft} rate between the version
of Daala presented in~\cite{EggePCS} and the current version on
the \emph{subset1} test set. Lower is better. \label{tab:bdrate-pcs-badchroma} }
\end{table}

\begin{table}
\centering{%
\begin{tabular}{|c|c|c|c|}
\hline 
Metric & Low (\%) & Medium (\%) & High (\%)\tabularnewline
\hline 
\hline 
PSNR & -3.6 & -3.0 & -1.7\tabularnewline
\hline 
PSNR-HVS & -2.9 & -2.2 & -0.9\tabularnewline
\hline 
SSIM & -1.5 & -1.6 & -1.0\tabularnewline
\hline 
FAST-SSIM & +2.7 & +3.4 & +3.4\tabularnewline
\hline 
\end{tabular}}\caption{Bjøntegaard-delta rate impact of the deringing filter on the \emph{subset1}
test set. Lower is better.\label{tab:bdrate-deringing}}
\end{table}

\begin{figure*}
\centering{\includegraphics[width=0.39\textwidth]{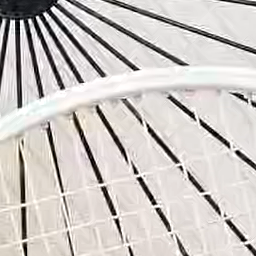}\hspace{0.4cm}\includegraphics[width=0.39\textwidth]{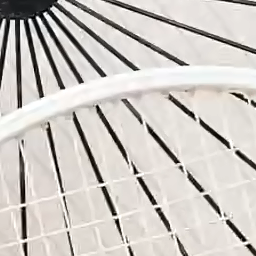}}\caption{Compressing the ``bike'' image at 0.25 bit/pixel. Left: Previous
results presented in~\cite{pcs2015}. Right: current result.\label{fig:Compressing-the-bike-image}}
\end{figure*}

Fig.~\ref{fig:Compressing-the-bike-image} shows the visual improvement
on the picture that was the most problematic~\cite{DaedeDCC} for
Daala in~\cite{pcs2015}. Both the reduction in deringing (due to
deringing and reduced lapping) and the reduction in chroma artifacts
(due to finer chroma quantization) are visible. 

\begin{figure}
\centering{\includegraphics[width=0.9\columnwidth]{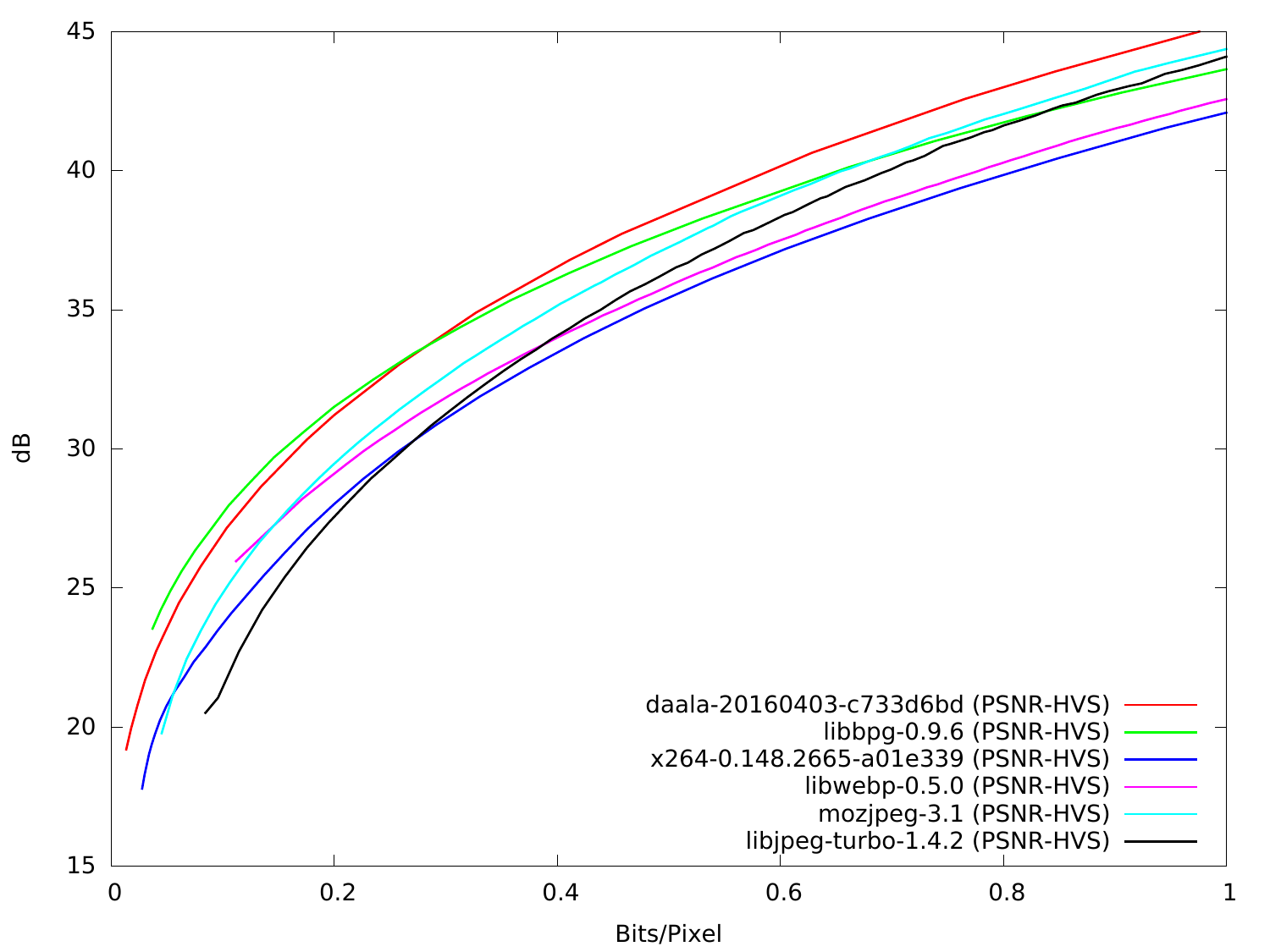}}\caption{Rate-distortion comparison between different codecs using PSNR-HVS
on \emph{subset1}.\label{fig:RD-curve-PSNR-HVS}}
\end{figure}

Fig.~\ref{fig:RD-curve-PSNR-HVS} shows the PSNR-HVS rate distortion
performance of Daala compared to JPEG (libjpeg-turbo~\cite{libjpegWebsite}
and mozjpeg~\cite{mozjpegWebsite}), WebP~\cite{WebPWebsite}, x264~\cite{x264Website},
and BPG (HEVC)~\cite{BPGWebsite}, using PSNR-HVS for \emph{subset1}.
The default compression parameters were used with all codecs with
the exception of x264 which used the command line: {\tt --preset placebo --crf=}.
Subjective results from the Image Compression Grand Challenge at ICIP
2016 will provide more information on the perceptual quality of Daala
as compared to other codecs. 

\section{Future Work}

The Daala bit-stream has not been frozen yet, so it is still being
improved. Among the recent features not demonstrated here are support
for image bit depths up to 12 bits, as well as a special coding tool
for non-photographic computer-generated images~\cite{valinL1TW},
which are generally better compressed with pixel prediction than block-based
transform coding. However, some features such as metadata support
and lossless RGB are still missing.

\bibliographystyle{IEEEbib}
\bibliography{daala}

\begin{thebibliography}{10}

\bibitem{EggePCS}
N.~E. Egge, J.-M. Valin, T.~B. Terriberry, T.~Daede, and C.~Montgomery,
\newblock ``Using {Daala} intra frames for still picture coding,''
\newblock in {\em Proceedings of Picture Coding Symposium}, 2015.

\bibitem{pcs2015}
``{Picture Coding Symposium} 2015 feature event: Evaluation of current and
  future image compression technologies,'' \url{http://www.pcs2015.org/}.

\bibitem{MalvarS89}
Henrique~S. Malvar and David~H. Staelin,
\newblock ``The {LOT}: Transform coding without blocking effects.,''
\newblock {\em IEEE Trans. Acoustics, Speech, and Signal Processing}, vol. 37,
  no. 4, pp. 553--559, 1989.

\bibitem{Tran2003}
T.D. Tran, Jie Liang, and Chengjie Tu,
\newblock ``Lapped transform via time-domain pre- and post-filtering,''
\newblock {\em Signal Processing, IEEE Transactions on}, vol. 51, no. 6, pp.
  1557--1571, June 2003.

\bibitem{DaedeDCC}
T.~J. Daede, N.~E. Egge, J.-M. Valin, G.~Martres, and T.~B. Terriberry,
\newblock ``{Daala}: A perceptually-driven next generation video codec,''
  \texttt{arXiv:1603.03129 [cs.MM]} \url{http://arxiv.org/abs/1603.03129},
  2016.

\bibitem{derfTools}
T.~B. Terriberry,
\newblock ``Coding tools for a next generation video codec,''
  \url{https://tools.ietf.org/html/draft-terriberry-codingtools-02}, 2015.

\bibitem{valin2015spie}
J.-M. Valin and T.~B. Terriberry,
\newblock ``Perceptual vector quantization for video coding,''
\newblock in {\em Proceedings of SPIE Visual Information Processing and
  Communication}, 2015, vol. 9410, pp. 941009--941009--11,
\newblock \texttt{arXiv:1602.05209 [cs.MM]}
  \url{http://arxiv.org/abs/1602.05209}.

\bibitem{egge2015spie}
N.~E. Egge. and J.-M. Valin,
\newblock ``Predicting chroma from luma with frequency domain intra
  prediction,''
\newblock in {\em Proceedings of SPIE Visual Information Processing and
  Communication}, 2015, vol. 9410, pp. 941008--941008--10,
\newblock \texttt{arXiv:1603.03482 [cs.MM]}
  \url{http://arxiv.org/abs/1603.03482}.

\bibitem{ValinDeringing}
J.-M. Valin,
\newblock ``The {Daala} directional deringing filter,''
  \texttt{arXiv:1602.05975 [cs.MM]} \url{http://arxiv.org/abs/1602.05975},
  2016.

\bibitem{deringing-demo}
J.-M. Valin,
\newblock ``A deringing filter for daala... and beyond,''
  \url{https://people.xiph.org/~jm/daala/deringing_demo/}, 2016.

\bibitem{stuiver1998piecewise}
Lang Stuiver and Alistair Moffat,
\newblock ``Piecewise integer mapping for arithmetic coding,''
\newblock in {\em Data Compression Conference, 1998. DCC'98. Proceedings}.
  IEEE, 1998, pp. 3--12.

\bibitem{PSNRHVSM}
N.~Ponomarenko, F.~Silvestri, K.Egiazarian, M.~Carli, and V.~Lukin,
\newblock ``On between-coefficient contrast masking of dct basis functions,''
\newblock in {\em Proceedings of Third International Workshop on Video
  Processing and Quality Metrics for Consumer Electronics VPQM-07}, 2007.

\bibitem{FastSSIM}
Ming-Jun Chen and Alan~C. Bovik,
\newblock ``Fast structural similarity index algorithm,''
\newblock {\em Journal of Real-Time Image Processing}, vol. 6, no. 4, pp.
  281--287, Dec. 2011.

\bibitem{testing-draft}
J.~Moffitt T.~Daede,
\newblock ``Video codec testing and quality measurement,''
  \url{https://tools.ietf.org/html/draft-daede-netvc-testing}, 2015.

\bibitem{libjpegWebsite}
``libjpeg-turbo website,'' \url{http://libjpeg-turbo.org/}.

\bibitem{mozjpegWebsite}
``mozjpeg website,'' \url{https://github.com/mozilla/mozjpeg}.

\bibitem{WebPWebsite}
``Web{P} website,'' \url{https://developers.google.com/speed/webp/}.

\bibitem{x264Website}
``x264 website,'' \url{http://www.videolan.org/developers/x264.html}.

\bibitem{BPGWebsite}
``{BPG} website,'' \url{http://bellard.org/bpg/}.

\bibitem{valinL1TW}
Jean-Marc Valin,
\newblock ``Screencasting considerations and {L1}-tree wavelet coding,''
  \url{https://tools.ietf.org/html/draft-valin-netvc-l1tw-01}, 2015.

\end{thebibliography}

\end{document}